\documentclass{Martin}

\hypersetup{
    colorlinks,
    linkcolor={red!50!black},
    citecolor={blue!50!black},
    urlcolor={blue!80!black}
}

\usepackage[bitstream-charter]{mathdesign}
\DeclareSymbolFont{usualmathcal}{OMS}{cmsy}{m}{n}
\DeclareSymbolFontAlphabet{\mathcal}{usualmathcal}

\usepackage{lineno} 
\usepackage{graphicx,epstopdf,color} 
\usepackage{amsfonts}
\usepackage{amsmath,mathrsfs}
\usepackage{bm}
\usepackage{ucs} 
\usepackage[T1]{fontenc} 
\usepackage{lmodern} 
\usepackage{microtype} 
\usepackage[final,inline]{showlabels} 
\usepackage{mdframed,enumitem}
\usepackage{ifpdf}

\def\tr{\text{tr}}

\newcommand{\ev}[1]{\left\langle #1 \right\rangle}
\newcommand{\bra}[1]{\left\langle#1\right|}
\newcommand{\ket}[1]{|#1\rangle}

\newcommand{\norm}[1]{\left\Vert #1\right\Vert}

\newcommand{\kett}[2]{\ket{#1}\otimes\ket{#2}}

\newcommand{\kettttt}[5]{\ket{#1}\otimes\ket{#2}\otimes\ket{#3}
  \otimes\ket{#4}\otimes\ket{#5}}
\newcommand{\ketABCD}[4]{\ket{#1}_{\!\A}\otimes\ket{#2}_\B
  \otimes\ket{#3}_\C\otimes\ket{#4}_\D}

\renewcommand{\i}{\text{i}}
\newcommand{\f}{\text{f}}

\newcommand{\z}{\text{z}}

\newcommand{\up}{\uparrow}
\newcommand{\dw}{\downarrow}

\renewcommand{\b}{\beta}

\newcommand{\A}{\text{A}}
\newcommand{\B}{\text{B}}
\newcommand{\D}{\text{D}}
\newcommand{\R}{\text{R}}
\newcommand{\bR}{\overline{\text{R}}}
\newcommand{\M}{\text{M}}
\newcommand{\bM}{\overline{\text{M}}}
\newcommand{\C}{\text{C}}
\newcommand{\bC}{\overline{\text{C}}}
\newcommand{\W}{\text{W}}
\newcommand{\EMO}{\emph{ensemble of macroscopic objects}}
\renewcommand{\EMO}{EMO}
\newcommand{\ETH}{\emph{eigenstate thermalization hypothesis}}
\renewcommand{\ETH}{ETH}
\newcommand{\AiP}{``{All is $\Psi$}''}

\renewcommand{\not}[1]{}
\newcommand{\cancel}[1]{\textcolor{red}{#1}}
\renewcommand{\cancel}[1]{}

\def\ie{{i.e.},\ }

\def\etal{{et al.}}

\def\cf{{cf.}\ }

\begin{document}

\begin{center}{\Large \textbf{
Interlinking and the Emergence of Classical Physics\\ in Quantum Theory\\
}}\end{center}

\begin{center}
Martin Greiter
\end{center}

\begin{center}
Institute for Theoretical Physics, University of
  Würzburg, 
97074 Würzburg, Germany\\
{\small \sf greiter@physik.uni-wuerzburg.de}
\end{center}

\begin{center}
\today
\end{center}


\section*{Abstract} {\bf The main argument by proponents of Many-World interpretations of quantum mechanics is that as more and more previously disentangled degrees of freedom become entangled with the microscopic degree we measure, there is no way of telling when the measurement (in the sense of a collapse of the wave function) should occur. Here, we introduce the concept of quantum interlinking, and argue that all macroscopic objects in the universe are connected through links of mutual entanglement while the objects themselves are (for the most part) not entangled.  The measurement occurs when the degree we measure becomes entangled with any macroscopic object, and hence interlinked with all of them.  This picture resolves long standing paradoxes such as Schrödinger's cat and
EPR.} 

\vspace{10pt}
\noindent\rule{\textwidth}{1pt}
\tableofcontents\thispagestyle{empty} 
\vspace{10pt}

\section{Introduction}
\label{introduction}

The subject of this article is the emergence of classical physics in quantum theory, which prominently includes the process of measuring quantum mechanical degrees of freedom with classical devices\cite{bohr28n580,
schroedinger35naturwissenschaften807,
everett57rmp454, 
mermin85pt38,
zurek02proc,  
coleman2011.12671prsty,
leggett05s871,
SaundersBarrettKentWallace10,%
mermin14n421,
bassi-13rmp471,
arndt-14np271
}.
Before diving into the details, however, I wish to engage in two digressions. First, I wish expand on the abstract by briefly explaining what quantum interlinking is.  Imagine three (macroscopic) systems A, B, and C, and assume that A is entangled with B, B is entangled with C, but A and C are neither entangled nor share mutual information\cite{NielsenChuang10}.  We will show below that this is possible.  Classically, A and C are as independent as they could be.  In quantum theory, however, they are interlinked, which means that we cannot have a complete description of A which does not also describe C, since the overall wave function describing A, B, and C cannot be factorised.  We will see below that if we assume a wave function for the universe, it will factorize into a very large number of independent wave functions describing microscopic degrees of freedom like electrons in filled shells, and \emph{one single, gigantic wave function describing all the interlinked, macroscopic objects.}  This simple observation has far-reaching consequences for the interpretation of our classical experience.

Second, I wish to mention why I have been interested in this topic after decades of sharing Feynman's pragmatic view\cite{feynman82ijtp467} on whether it is fruitful to think about the interpretation of quantum mechanics.  The fundamental question I have always been interested in is whether there is a fundamental ``force'' in nature towards (or reason for) the development of structures, and life in particular.  The principles of both classical and quantum dynamics we teach in physics curriculums are deterministic.  (Prigogine\cite{Prigogine80} has advocated that the origin of structures may be found in the second law of thermodynamics, but it does not appear plausible how a statistical theory without dynamics could account for the origin of life.)  The only place where something non-deterministic appears to happen is in between, when a measurement occurs.  Therefore, it seems reasonable to start by thinking about the measurement process.  I will return to this point at the end.

To establish notation and terminology, I will begin with a brief review of what is generally agreed on in quantum mechanics\cite{Gottfried66}:
\begin{enumerate}[label=(\alph*)]
\item At any fixed time, the state of a system is described by a (normalized) vector in Hilbert space, $\ket{\psi}$.

\item The state vector evolves according to Schrödinger's equation,
\begin{linenomath}\begin{align}
   \label{eq:schroedinger}
    \i \frac{\partial}{\partial t}\ket{\psi}= H \ket{\psi},   
\end{align}\end{linenomath}
where the Hamilton $H$ of the system is a linear, self-adjoint operator.

\item Observables are likewise described by linear, self-adjoint operators.  If $\ket{\psi}$ is in an eigenstate of an observable $A$,
$A\ket{\psi}=a\ket{\psi}$, the observed value of $A$ is $a$.

\item\label{projection} Every measurement of $A$ with a device described by classical physics yields one of the eigenvalues of $A$.  The probability of finding a particular eigenvalue $a$ when measuring $\ket{\psi}$ is $\norm{P(A,a)\ket{\psi}}^2$, where $P(A,a)$ is the projection operator on the subspace of states with eigenvalue $a$ (Born's rule). If $A$ is measured again thereafter, the observed value will 
be $a$ again.
\end{enumerate}

According to the Copenhagen interpretation\cite{bohr28n580} of 1927, \ref{projection} implies that the state of the system after the first measurement is given by
  \begin{linenomath}\begin{align}
    \label{eq:projection} 
    \frac{P(A,a)\ket{\psi}}{\norm{P(A,a)\ket{\psi}}}.
  \end{align}\end{linenomath}
This is referred to as the ``collapse'' of the wave function.  The probability in the process is assumed to be frequentist, \ie due to the occurence of random events, as opposed to a Bayesian probability, which is a subjective probability an observer assigns due to inaccessibility of information.

\section{Problems with the Copenhagen interpretation}
\label{copenhagen}

There are two reasons why this interpretation
is problematic.  The first is that the laws of quantum mechanics, and in particular the evolution according to \eqref{eq:schroedinger}, have been tested and verified to highest precision for almost a century\cite{arndt-14np271}.  Since this formalism is deterministic, it cannot describe a collapse of a wave function along the lines of \eqref{eq:projection}, and all attempts to extend it in a way consistent with the entirety of experimental observations have been inconclusive\cite{ghirardi-86prd470,diosi87pla377,penrose96grg581,bassi-13rmp471,arndt-14np271,
donadi-21np74}.  
A related issue is that in the absence of a viable formalism, we have little guidance where to place the boundary between quantum and the classical domains, as illustrated by Schrödinger's cat\cite{schroedinger35naturwissenschaften807} or Wigner's friend\cite{Wigner61,bong-20np271}.

The second problem is that the assignment of frequentist probabilities appears to violate the principle of locality in an Einstein--Podolski--Rosen\cite{einstein-35pr777,reid-09rmp1727} (EPR) setting.  I will elaborate this point now.

%
%

In an EPR setting, one considers a pure quantum state of two entangled degrees of freedom, moves them far apart, and measures observables which reflect the correlation due to the entanglement in space-like separated  regions of spacetime $\R$ and $\bR$.  Then causality precludes any interference between the measurements.  According to the Copenhagen interpretation, each measurement will cause a collapse of the wave function, with a frequentist probability assigned to each.  Since the results are correlated, however, both cannot be random.  Note further that there is no objective sense as to which measurement occurs first, since this depends on the frame of reference.

For definiteness, let the pure state we prepare be a spin singlet\cite{Bohm51} of two spin half particles 1 and 2,
\begin{linenomath}\begin{align}
  \label{eq:epr}
  \ket{\psi}=\frac{1}{\sqrt{2}}
                    \Bigl(\,\ket{\!\up_1}\otimes\,\ket{\!\dw_2}
                    -\,\ket{\!\dw_1}\otimes\,\ket{\!\up_2}\Bigr),
\end{align}\end{linenomath}
where $\up_{i}$ and $\dw_{i}$ refer to the $\sigma^\z$ eigenvalues $+1$ and $-1$ of particle $i$, respectively.  As one measures $\sigma^\z_{(1)}$ of particle 1 in region $\R$ and $\sigma^\z_{(2)}$ of particle 2 in region $\bR$, and later compares the results, one will find that the product of the eigenvalues is always $-1$, \ie that one of the spins is $\up$ and the other $\dw$.  The outcome of the ``first'' measurement determines the outcome of the ``second'', and we can assign a frequentist probability to at most one of them.  

This problem led Einstein \etal\cite{einstein-35pr777} to conclude in 1935 that the theory of quantum mechanics cannot be complete in the sense that additional information must be available in regions $\R$ and $\bR$ (so called ``hidden variables'').  This possibility was subsequently ruled out by experiment\cite{%
aspect-82prl1804, 
pan-00n515, 
reid-09rmp1727,hensen-15n682,giustina-15prl250401}.
A pedagogically outstanding review of how this can be done in principle has been given by Coleman\cite{coleman2011.12671prsty}, based on Gedanken experiments by Bell\cite{bell64physics195} and by Greenberger, Horne, and Zeilinger\cite{Greenberger-89proc}.  The most significant aspect of EPR is that it cannot be reconciled with locally induced collapse models\cite{ghirardi-86prd470,diosi87pla377,penrose96grg581,bassi-13rmp471,arndt-14np271,donadi-21np74}.

%
%

\section{Many-Worlds interpretations}
\label{MWI}

Both these problems may be resolved by turning to Many-Worlds interpretations\cite{
everett57rmp454,DeWittGraham73,
coleman2011.12671prsty, 
SaundersBarrettKentWallace10,
Wallace12,
vaidman16jpcs012020,vaidman18coll%
} (MWIs).  I will limit the presentation here to what I will need to reconcile Many-World interpretations with the Copenhagen interpretation further below. 

The idea is simply that there is only quantum mechanics (
``{All is $\Psi$}'', in the words of Vaidman\cite{vaidman16jpcs012020}), and only deterministic evolution according to Schrödinger's equation \eqref{eq:schroedinger}. 
Measurements do not entail a projection or collaps of the wave function.  The definiteness of our daily life experience is subjective only.  All probabilities are Bayesian. 

To be concise, consider a spin half particle in a superposition of eigenstates of $\sigma^z$,
\begin{linenomath}\begin{align}
  \label{eq:mw}
  \ket{\psi}=u\,\ket{\!\up}+v\,\ket{\!\dw},
\end{align}\end{linenomath}
where $u$ and $v$ are complex coefficients  such that $\ket{\psi}$ is normalized.  Consider further a $\sigma^z$ measuring device M with Hilbert space $\M=\{\ket{\M_0}, \ket{\M_\up}, \ket{\M_\dw}\}$, and an observer consciousness C with Hilbert space $\C=\{\ket{\C_0}, \ket{\C_\up}, \ket{\C_\dw}\}$.  The subscript $0$ indicates states where no measurement has taken place,
while $\up$ and $\dw$ indicate states where $\sigma^z$ eigenvalues 
$+1$ or $-1$ have been measured or perceived. According to MWIs, a measurement of \eqref{eq:mw} by the observer will evolve the initial state
\begin{linenomath}\begin{align}
  \label{eq:mw_i}
  \ket{\psi_\i}=\Bigl( u\,\ket{\!\up}+v\,\ket{\!\dw} \Bigr)
  \otimes\kett{\M_0}{\C_0}
\end{align}\end{linenomath}
following the von Neumann chain\cite{Neumann55} via the intermediate state
\begin{linenomath}\begin{align}
  \label{eq:mw_fM}
  \ket{\psi_{\M}}
  =\Bigl( u\,\ket{\!\up}\otimes\ket{\M_\up}+v\,\ket{\!\dw}\otimes\ket{\M_\dw}\Bigr)
  \otimes\ket{\C_0}
\end{align}\end{linenomath}
into the final state
\begin{linenomath}\begin{align}
  \label{eq:mw_fMC}
  \ket{\psi_{\M,\C}}
  =u\,\ket{\!\up}\otimes\kett{\M_\up}{\C_\up}
                    +v\,\ket{\!\dw}\otimes\kett{\M_\dw}{\C_\dw}
\end{align}\end{linenomath}
without a projection onto one of the terms on the left of \eqref{eq:mw_fMC}.
If we define a definiteness operator $D$\cite{Albert94,coleman2011.12671prsty} such that it will yield eigenvalue $+1$ if the observer perceives a definite outcome, 
\begin{linenomath}\begin{align}
  \label{eq:mw_D}
  D\ket{\C_\up}=\ket{\C_\up},\ 
  D\ket{\C_\dw}=\ket{\C_\dw},
\end{align}\end{linenomath}
and annihilate all states orthogonal to those, the final state \eqref{eq:mw_fMC} will trivially be an eigenstate of $D$ with eigenvalue $+1$ as well.  In other words, the state will be perceived as definite
even if it is not projected
onto an eigenstate of $\sigma^z$.

The evolution of the state, of course, does not stop with $\eqref{eq:mw_fM}$ or $\eqref{eq:mw_fMC}$, as the state will rapidly become entangled with other degrees of freedom.  One can now take the view of a local Hilbert space which is coupled weakly to an external bath.  This causes decoherence\cite{zeh70fp69,zurek81prd1516,
zurek02proc,
JoosZehKieferGiuliniKupschStamatescu03,zurek03rmp715,
schlosshauer05rmp1267,
schlosshauer19pr1}
and necessitates that the local description will be
in terms of a density matrix rather than a state vector.  
The view taken in MWIs, however, is that as more and more degrees of freedom become entangled with the two orthogonal amplitudes in \eqref{eq:mw_i}, the state will remain in a coherent superposition of both, until we end up in a coherent superposition of two ``worlds''.  
Regardless of which view is taken,
the preferred basis for these two worlds ($\up$ and $\dw$ in the $\sigma^z$ basis in our example) is selected via decoherence\cite{saunders93fp1553,SaundersBarrettKentWallace10}.  Formally, we may introduce a Hilbert space of the world $\W=\{\ket{\W_0}, \ket{\W_\up}, \ket{\W_\dw}\}$ and write the final state after the observer became entangled as
\begin{linenomath}\begin{align}
  \label{eq:mw_fMCW}
  \ket{\psi_\f}
  =u\,\ket{\!\up,\M_\up,\C_\up,\W_\up}+v\,\ket{\!\dw,\M_\dw,\C_\dw,\W_\dw}.
\end{align}\end{linenomath}
If an observer has the conscious perception of having measured an $\up$ spin, there exists another world where the (same) observer has the conscious perception of having measured an $\dw$ spin.  (I have put ``same'' in brackets because they are only identical up to the time of the measurement.)  In general, each measurement of a state which is not an eigenstate of the observable we measure yields a branching into different worlds.  
%
%
%
The (Bayesian) probability for any observer to find himself in the branch where the measured value of observable $A$ is $a$ is assumed
(and has been derived under certain assumptions) to be $\norm{P(A,a)\ket{\psi}}^2$ (\cf point \ref{projection} in the Introduction)\cite{
SaundersBarrettKentWallace10, vaidman12proc,sebens-18bjps25,kent15fp211,vaidman20proc}.


The MWIs clearly resolve the first problem of the Copenhagen interpretation.  Now consider measurements of EPR states in MWIs.  Including Hilbert spaces for $\sigma_\z$ measuring devices M and $\bM$ and observer consciousnesses C and $\bC$ in spacetime regions $\R$ and $\bR$, respectively, we write the initial state as
\begin{linenomath}\begin{align}
 \label{eq:epr_i}
  \ket{\psi_\i}=\frac{1}{\sqrt{2}}
                    \Bigl(\,\ket{\!\up_1}\otimes\,\ket{\!\dw_2}
                    -\,\ket{\!\dw_1}\otimes\,\ket{\!\up_2}\Bigr)
  \otimes\kettttt{\M_0}{\C_0}{\bM_0}{\bC_0}{\W_0}
    \hspace{2pt}
\end{align}\end{linenomath}
where we take the Hilbert space of the world to be 
$$\W=\bigl\{\ket{\W_0}, \ket{\W_{\up_1\up_2}}, \ket{\W_{\up_1\dw_2}}, \ket{\W_{\dw_1\up_2}} , \ket{\W_{\dw_1\dw_2}} \bigr\}.$$  
%
%
When we carry out measurements in R and $\bR$, 
the final state will evolve following the von Neumann chain 
into
\begin{linenomath}\begin{align}
 \label{eq:epr_f}\nonumber
  \ket{\psi_\f}=\frac{1}{\sqrt{2}}
  \Bigl(\,&\ket{\!\up_1}\otimes\,\ket{\!\dw_2} 
   \otimes\kettttt{\M_\up}{\C_\up}{\bM_\dw}{\bC_\dw}{W_{\up_1\dw_2}}
  \hspace{10pt}\\[5pt] 
  &-\text{ same term with}\up\ \leftrightarrow\ \dw\, 
    \Bigr). 
\end{align}\end{linenomath}
This is a coherent superpositions of two worlds, which both have the property that the product of the eigenvalues obtained via measurements of $\sigma^\z_{(1)}$ and $\sigma^\z_{(2)}$ is $-1$.

\not{
Note that if only the observer in region R carries out the measurement,
leading to the intermediate state
\begin{linenomath}\begin{align}
 \label{eq:epr_f1}\nonumber
  \ket{\psi_\f}
  =\frac{1}{\sqrt{2}}\Bigl(
  &\,\ket{\!\up_1,\M_\up,\C_\up;\dw_2,\bM_0,\bC_0;W_{\up_1\dw_2}}\\
  -&\,\ket{\!\dw_1,\M_\dw,\C_\dw;\up_2,\bM_0,\bC_0;W_{\dw_1\up_2}}\Bigr),
\end{align}\end{linenomath}
and the second observer with consciousness $\bC$ later learns through knowledge of the set-up and 
communication with observer C what he would have obtained if he had measured $\sigma^\z_{(2)}$ while in region $\bR$, the state would be 
\begin{linenomath}\begin{align}
 \label{eq:epr_f2}\nonumber
  \ket{\psi_\f}
  =\frac{1}{\sqrt{2}}\Bigl(
  &\,\ket{\!\up_1,\M_\up,\C_\up;\dw_2,\bM_0,\bC_\dw;W_{\up_1\dw_2}}\\
  -&\,\ket{\!\dw_1,\M_\dw,\C_\dw;\up_2,\bM_0,\bC_\up;W_{\dw_1\up_2}}\Bigr).
\end{align}\end{linenomath}
If $\bC$ would now the device $\bM$ to measure the spin of particle 2 nontheless, the final state would again be given by \eqref{eq:epr_f}.
}

\cancel{The paragraph below could be omitted.}
We see that within MWIs, the space-like separation of the regions R and $\bR$, in which measurements M and $\bM$ take place, does not render the situation different from what it would be if we were to measure $\sigma^\z_{(1)}$ and $\sigma^\z_{(2)}$ of particles 1 and 2 directly and locally after preparing the singlet state \eqref{eq:epr}.  Note further that if the second observer $\bC$ obtains the information by asking C rather than by conducting the measurement $\bM$, the final state vector is given by \eqref{eq:epr_f} with $\bM_\dw$ and $\bM_\up$ replaced by $\bM_0$.  Therefore, it is irrelevant for the state of the second observer whether he measures the spin of particle 2 or asks the first observer.

In summary, the analysis presented so far suggests that MWIs are fully consistent, both internally and with the observed phenomenology, while the Copenhagen interpretation is not.  Still, a poll\cite{schlosshauer-13shpsb222} from 2013 showed that 42\% of all physicists subscribe to Copenhagen, while only 18\% subscribe to MWIs.  The reason is presumably that the notion of constant branchings into countless many worlds is deeply unappealing, if not altogether inadequate\cite{kent90ijmpa1745,
kent10proc
}.
No one appears to like it, but some of the greatest minds of our time subscribe to it because they are even less willing to accept logical inconsistencies.

%
%

\section{Interlinking and the ensemble of macroscopic objects}
\label{interlinking}

In this work I will show that, when consequently thought through, the assumption that there is only quantum mechanics 
 will not (necessarily) imply many worlds, but effectively lead to a ``classical reality'' onto which states become projected upon measurement, \ie to a phenomenology close to the Copenhagen interpretation, but without the inconsistencies described above.  This is not to say that I will be able to rule out MWIs---I think this would be impossible 
with the body of available experimental evidence---but
rather that I will show that there is no need to invoke them.

I will begin with a few assumptions.  Most of them can be relaxed later on, but it is helpful to start from a concise picture. 

\begin{enumerate}[label=(\roman*)]
\item\label{allispsi} The fundamental theory is a quantum theory.  The entire universe can be described by a solution of this quantum theory, which for simplicity we call wave function $\Psi$.

\item\label{linear} The evolution of $\Psi$ is, to an approximation we have not been able to challenge, given by the linear regime of the quantum theory. For simplicity, let us assume time is fundamental (as opposed to emerging) and let us refer to the theory describing this evolution as the Schrödinger equation.  For time to be meaningful, $\Psi$ must not be an eigenstate of the time evolution operator.

\item\label{bigbang} The universe started with the big bang, and at that time, many degrees of freedom of the universe were entangled with their environments.  We expect that there is still significant entanglement.

\item\label{oneworld} For simplicity, we further assume that at a time we refer to as the present, there is only one ``world''.  Since all the branchings into other ``worlds'', should they have occurred in the past, have no influence on our perception of the present, any consistent theory based on this assumption will be sufficient.
\end{enumerate}

The reality of our daily life experience, however, does not appear to be governed by a quantum theory or described by a pure state $\Psi$, as most of it is described by either classical dynamics or statistical physics.  The latter, however, can be understood as emergent.  With the known exception of systems which are either integrable or display single or many-body 
localization,
interacting quantum systems thermalize\cite{deutsch91pra2046,
srednicki94pre888,
srednicki99jpa1163,
rigol-08n854,
GemmerMichelMahler09,
gogolin-16rpp056001,
dalessio-16ap239,
deutsch18rpp082001,
garrison-18prx021026}, 
in the sense that if we partition a system in a pure quantum state $\ket{\psi(t)}$ into a small subsystem $A$ and a large environment $B$, the density matrix of the subsystem $A$ obtained by tracing out the environment $B$,
\begin{linenomath}\begin{align}\nonumber
  \rho_A(t)\equiv\tr_{B}\bigl(\ket{\psi(t)}\!\bra{\psi(t)}\bigr),
\end{align}\end{linenomath}
will converge towards a Boltzmann distribution, 
\begin{linenomath}\begin{align}\nonumber
  \rho_A(\b)\equiv\tr_{B}\bigl(\rho(\b)\bigr)\quad\text{with}\quad
  \rho(\b)=\frac{1}{Z}\exp\left(-\b H\right).
\end{align}\end{linenomath}
The statement becomes exact in the limit where both the size of $B$ and the time we wait is taken to infinity.  If $\ket{\psi(t)}$ is an eigenstate of the system with energy $E_\psi$,
the eigenstate thermalization hypothesis 
(ETH) of Deutsch\cite{deutsch91pra2046}
and Srednicki\cite{srednicki94pre888,
srednicki99jpa1163}
states that the equivalence $\rho_A(t)=\rho_A(\b)$ holds for small subsystems A which are weakly coupled to the environment B
, with the inverse temperature $\b$ determined by 
$\ev{H}_{\b}=E_\psi$.
Even though the ETH has so far not been shown to hold for interacting systems (without integrability or localization) in general, it is supported by a significant body of numerical evidence\cite{rigol-08n854,dalessio-16ap239,garrison-18prx021026}.  
An important consequence of quantum thermalization, and the ETH in particular, is that the entanglement entropy\cite{Neumann55,NielsenChuang10} 
of a subsystem is equal to the thermal entropy.

The ``{All is $\Psi$}'' assumption \ref{allispsi} hence implies that \emph{all of entropy is entanglement entropy}. 
The second law of thermodynamics merely states that the entanglement of subsystems with its environment does not decrease under usual circumstances.

%
%

We now identify the classical reality we perceive as such and describe by classical dynamics with the \emph{ensemble of macroscopic objects} (EMO), in which all objects are connected through chains of entangled links.  This does not imply that all objects are mutually entangled or share mutual information, but merely that all the individual degrees of freedom are entangled with other degrees of freedom in their environment, which are then again entangled with more degrees of freedom, and so on.  In this way, chains of entangled links connect all degrees of freedom in the \EMO .

To illustrate this concept of \emph{quantum interlinking}, consider a simple example consisting of four qubits (two-state systems) $\A,\B,\C,\D$ with states $$\ket{n_1;n_2;n_3;n_4}\equiv\ketABCD{n_1}{n_2}{n_3}{n_4},$$ where the $n_i$'s
can take the values $0$ or $1$. 
Now consider the state 
\begin{linenomath}\begin{align}
  \label{eq:psiabc}
  \ket{\psi}
  &= \frac{1}{2}\sum_{i=0}^1 \sum_{k=0}^1
    \ket{i;(i+k)\hspace{-6pt}\mod 2;k;k}\\[2pt]
  &=\frac{1}{2}\Bigl(\ket{0;0;0;0}+\ket{0;1;1;1}
    +\ket{1;1;0;0}+\ket{1;0;1;1} \Bigr).\nonumber
\end{align}\end{linenomath}
With the entanglement entropy 
of system $\A$ given by
\begin{linenomath}\begin{align}\nonumber
  S(\A)= -\tr_\A(\rho_\A \ln \rho_\A)\quad \text{with}\
  \rho_\A=\tr_{\B\C\D}\bigl(\ket{\psi}\!\bra{\psi}\bigr),
\end{align}\end{linenomath}
we find $S(\A)=S(\D)=\ln 2$, and 
$S(\A,\D)=2\ln2$,  
for the joint entropy $$S(\A,\D)=-\tr_{\A\D}(\rho_{\A\D} \ln \rho_{\A\D})$$ of $\A$ and $\D$.  $\A$ and $\D$ are neither correlated nor share mutual information\cite{NielsenChuang10}, 
\begin{linenomath}\begin{align}\nonumber
  S(\A\!:\!\D)\equiv S(\A)+S(\D)-S(\A,\D)=0.  
\end{align}\end{linenomath}
While $\A$ and $\D$ in \eqref{eq:psiabc} are not entangled, they are still interlinked though the chain $\A\B\C\D$ of mutually entangled degrees of freedom.  It is not possible to factorize $\ket{\psi}$ into a product state, and hence not possible to describe $\A$ or $\D$ independently.  This simple example illustrates that (sub-)systems can be interlinked while they are neither entangled nor correlated.  There are no classical effects associated with quantum interlinking.


To quantify the interlinking of two (uncorrelated) subsystems $\A$ and $\A'$, one may consider the entanglement entropy for all possible ways to bisect the system such that $\A$ and $\A'$ belong to different sectors.  A suggestive measure for the interlinking $I(\A,\A')$ is then the minimal value of the entropy attainable.  In state \eqref{eq:psiabc}, $I(\A,\D)=\ln 2$, since $S(\A)=S(\A,\B)=$ 
$S(\A,\B,\C)=\ln 2$.  For a given macroscopic object, the amount of interlinking with the EMO is simply given by the thermal entropy of this object, since this entropy corresponds to the entanglement entropy between the object and the \EMO .

The important point in the present context is that unless it is possible to partition a macroscopic system such that the parts are not entangled, all the degrees of freedom are interlinked.  \emph{In a quantum theory, interlinked degrees of freedom cannot be described independently, even if they are classically independent.}  The assumption that all common macroscopic objects in the universe are interlinked is backed by the applicability of statistical mechanics and thermodynamics.  If all of entropy is interpreted as entanglement entropy, any object which carries entropy, and hence any object to which we can assign a (non-zero) temperature, is entangled with its environment, and hence part of the \EMO .  On the other hand, if we prepare an isolated system in a pure quantum state, it will not be possible to assign a temperature to it.  In practice, of course, macroscopic systems become very rapidly entangled with the environment.

%
%

\section{The measurement process, Schrödinger's cat, and EPR}
\label{measurement}

The notion of the \EMO , and in particular quantum interlinking, is key to understanding the measurement process, or in general, the transition from a quantum to a classical description.  \emph{A measurement occurs when one or several microscopic degrees of freedom, which were previously disentangled from the \EMO , become interlinked with it through entanglement with degrees of freedom belonging to the \EMO .}  In the Copenhagen interpretation, this process 
corresponds to the collapse 
of the wave function.  To be consistent with a century of literature on the subject, we refer to this process as a ``measurement''.  Strictly speaking, however, this is not fully accurate as it is irrelevant whether we extract information or not.  For the transition from a quantum to a classical description, only interlinking is required.

The process is hence different from the chain envisioned by von Neumann\cite{Neumann55}, which we described by the sequence of states~\eqref{eq:mw_i}, \eqref{eq:mw_fM}, \eqref{eq:mw_fMC}, and \eqref{eq:mw_fMCW} in the example above.  While the correlations of the device M and the observer consciousness C with the initial spin \eqref{eq:mw} develop step by step, the interlinking occurs all at once.  Since M, C, and W are interlinked already, it is not possible to factorize them into a product state.  The correct way to write the initial state is hence  
\begin{linenomath}\begin{align}
  \label{eq:martin_i}
  \ket{\psi_\i}=\Bigl( u\,\ket{\!\up}+v\,\ket{\!\dw} \Bigr)
  \otimes\ket{\M_0,\C_0,\W_0}
\end{align}\end{linenomath}
rather than~\eqref{eq:mw_i}, which then evolves through  
\begin{linenomath}\begin{align}
  \label{eq:martin_fM}\nonumber
  \ket{\psi_{\M}}
  =u\,\ket{\!\up}\otimes\ket{\M_\up,\C_0,\W_\up}
  +v\,\ket{\!\dw}\otimes\ket{\M_\dw,\C_0,\W_\dw}\hspace{20pt}\\[-15pt]
\end{align}\end{linenomath}
into 
\begin{linenomath}\begin{align}
  \label{eq:martin_fMCW}\nonumber
  \ket{\psi_\f}
  =u\,\ket{\!\up}\otimes\ket{\M_\up,\C_\up,\W_\up}
  +v\,\ket{\!\dw}\otimes\ket{\M_\dw,\C_\dw,\W_\dw}.\hspace{20pt}\\[-15pt]
\end{align}\end{linenomath}
Note that through interlinking, the bifurcation into ``two worlds'' occurs already when the initial spin state \eqref{eq:mw} becomes entangled with M.


The first step \eqref{eq:mw_fM} in the von Neumann chain could only happen if the measuring device M was disentangled from everything else beforehand, which, as explained above, 
would require it to be in a pure quantum state and preclude the possibility to assign a temperature to it.


Since Wigner's friend\cite{Wigner61}, Schrödinger's cat\cite{schroedinger35naturwissenschaften807}, and the Geiger counter used in the thought experiments all have a temperature, they are all part of the \EMO .  The measurement occurs whenever the quantum mechanical degree of freedom subject to measurement becomes entangled with one of them.
In these settings, the Geiger counter becomes entangled first.  
Adding a cat, a friend, or walls around either will make no difference.

The EPR paradox is likewise resolved by taking into account that the measuring devices M and $\bM$ in regions R and $\bR$ are both part of the \EMO , and are as such interlinked.  The measurement takes place when one of the two mutually entangled EPR spins becomes entangled with either measuring device, since this automatically interlinks both spins with the \EMO .  Expressed in equations, if we measure spin 1 first, the initial state
\begin{linenomath}\begin{align}
 \label{eq:martin_epr_i}
  \ket{\psi_\i}=\frac{1}{\sqrt{2}}
  \Bigl(\,\ket{\!\up_1}\otimes\,\ket{\!\dw_2}-\,\ket{\!\dw_1}\otimes\,\ket{\!\up_2}\Bigr)
                    \otimes\ket{\M_0,\C_0,\bM_0,\bC_0,\W_0}
\end{align}\end{linenomath}
evolves into  
\begin{linenomath}\begin{align}
 \label{eq:martin_epr_fm1}\nonumber
  \ket{\psi_\f}
  =\frac{1}{\sqrt{2}}\Bigl(
  &\,\ket{\!\up_1}\otimes\,\ket{\!\dw_2}\otimes
    \ket{\M_\up,\C_0,\bM_0,\bC_0,W_{\up_1\dw_2}}\\
  -&\,\ket{\!\dw_1}\otimes\,\ket{\!\up_2}\otimes
     \ket{\M_\dw,\C_0,\bM_0,\bC_0,W_{\dw_1\up_2}}\Bigr) .
\end{align}\end{linenomath}
When we measure the second spin, device and spin are interlinked already, and the outcome is certain.  

%
%

\section{Gravity and non-linearities}
\label{gravity}

\cancel{The paragraphs below can be compressed into one.}  It is 
 more subtle to address the question whether both amplitudes in Eqs.~\eqref{eq:martin_fM} or \eqref{eq:martin_epr_fm1} persist, as MWIs would suggest, or collapse onto one of them, as suggested by the Copenhagen interpretation.  As mentioned above, the argument cited by proponents of MWIs is that there is nothing in the formalism to describe a collapse.  It would require the linear Schrödinger equation, 
or its relativistic generalizations, to be augmented by some non-linearity, and there is no experimental evidence for any such augmentation as larger and larger systems can be prepared in well-defined quantum states.  If a non-linearity, possibly in the spirit of Ghirardi, Rimini, and Weber\cite{ghirardi-86prd470}, or Diósi\cite{diosi87pla377} and Penrose\cite{penrose96grg581,penrose14fp557}, were to exist\cite{bassi-13rmp471}, so the line of argument, we would have noted it already\cite{arndt-14np271,donadi-21np74}.  Therefore, we have no choice but to adjust our interpretation to the equations we have, and hence to embrace MWIs.

An alternative way to present this dilemma is through the search for a boundary between quantum and classical domains, as illustrated so charmingly in a cartoon by Zurek\cite{zurek02proc}. 
In the cartoon, the horizontal axis denotes the size (\# of atoms) of the system, with $1$ atom firmly in the quantum and $10^{23}$ atoms firmly in the classical domain.  We know that the quantum domain exists, and whenever we push towards the classical domain, 
we find no indications of a boundary.  Therefore, there is only a quantum domain, and our perceived definiteness of classical reality is subjective only, as elaborated with 
the ``definiteness operator'' defined in \eqref{eq:mw_D} above. 

The problem with this view is once again the implicit assumption of a von Neumann chain.  If it was technically feasible to prepare a system of $10^{23}$ atoms in a pure quantum state, disentangled form the \EMO ,
we should in principle be able to observe interference phenomena.  (
Practical limitations arise from the smallness of $\hbar$.)  On the other hand, if a system consisting of $10^2$ atoms is entangled with the \EMO , we will not be able to observe interference phenomena.
Since the number of baryons (protons and neutrons) in the visible universe is of the order of $10^{80}$, it is very plausible that all systems we will ever be able to prepare in a laboratory will evolve to the greatest precision we will ever be able to access according to linear equations, while the corrections to them become meaningful only at much larger scales.  This is significant as we are dealing with scales at which the effects of (quantum) gravity, which
do not necessarily share the linearity of
Schrödinger's equation, need to be taken into account.

Let us recapitulate what we understand so far. When we measure the spin $\sigma^\z$ of the initial state \eqref{eq:mw_i} above, we entangle it with its environment and hence interlink it with the \EMO .  According to the linear Schrödinger equation, the evolution will be given by
\begin{linenomath}\begin{align}
  \label{eq:mw_W}
  \ket{\psi_\i}&=\Bigl(u\,\ket{\!\up}+v\,\ket{\!\dw}\Bigr)\otimes\ket{W_0}
  \\[5pt]\nonumber
  &\to\ \ket{\psi_\f}=u\,\ket{\!\up}\otimes\ket{\W_\up}
    +v\,\ket{\!\dw}\otimes\ket{\W_\dw},\hspace{10pt}
\end{align}\end{linenomath}
where $\W_\up$ and $\W_\dw$ denote ``worlds'' in which the spin is measured and perceived $\up$ or $\dw$, respectively.  As 
in every scenario, decoherence is key, since it is responsible for the selection of the basis of these ``worlds'', \ie the $\sigma^\z$ basis in $\ket{\psi_\f}$ above.  The only reason to keep both amplitudes in $\ket{\psi_\f}$ is the linearity of Schrödinger's equation.  We have no reason to expect, however, that this linearity will prevail once gravity is included.  The ``worlds'' are given by the \EMO , which has a scale where gravity cannot possibly be neglected.  Therefore, MWIs are based on extrapolation of a set of equations to a regime where we have no reason to assume validity.  While Many Worlds seem inevitable when one thinks along the von Neumann chain, there is no need to invoke them in the framework I advocate here.

Let us now, for the sake of discussion, assume that a collapse occurs, due to non-linearities we have not yet been able to include in our equations describing the time evolution in quantum theories, and assign a frequentist probability to it.  Then the final state in \eqref{eq:mw_W} will evolve into one of the two amplitudes in the superposition, that is, into either 
$\,\ket{\!\up}\otimes\ket{\W_\up}$ or  $\,\ket{\!\dw}\otimes\ket{\W_\dw}$.  The spin, which was disentangled from the \EMO\ initially, became entangled and immediately disentangled again.  During the measurement process, it gained and lost entanglement entropy.  The evolution of the spin is adequately described by Schrödinger's 
equation before and after the measurement.

Note that the only assumption this framework depends on from the list above is the ``{All is $\Psi$}'' assumption \ref{allispsi}.  The (approximate) evolution according to a linear equation in \ref{linear} is experimentally verified and undisputed, and the entanglement within our universe in \ref{bigbang} follows through the \ETH\ from the applicability of statistical mechanics and thermodynamics.    Nothing depends on \ref{oneworld} anyway.  Note also that in \ref{allispsi}, we do not require that our universe is in a pure quantum state $\Psi$, since we can always purify
any mixed state through additional degrees of freedom we subsequently trace out\cite{NielsenChuang10}.  So the only assumption we have really made is that the fundamental theory is a quantum theory.


The resulting picture implies that when a spin is measured here on earth, the wave function will become interlinked instantly with stars on the other side of our galaxy.  Does this contradict the principle of relativity?  The answer is no, as no information is transmitted.  Interlinking happens in Hilbert space, not in real space, and has no classical or observable consequences.  When we calculate amplitudes in functional integrals, the choice of paths' we integrate over is likewise not constrained by the principle of relativity.  This appears to indicate that Hilbert space is fundamental, while the physical space subject to the principle of relativity is 
is an emergent (tensor product) structure within this space.
%
%
The possibility that space-time emerges from entanglement has recently been explored in anti-de Sitter space\cite{maldacena99ijtp1113,
ryu-06prl181602,
vanraamsdonk10grg2323,
evenbly-11jsp891,
swingle12prd065007,
maldacena-13fdp781,
vanraamsdonk20s198%
}.  

Another broad implication of the present proposal for the collapse of wave functions is that there cannot be a canonical quantization of gravity, as canonical quantization is inherently linear.  If the physical space we live in is emergent, however, we expect gravity to emerge with this space, and there is no need to quantize it\cite{sakharov00grg365, 
jacobson95prl1260,Volovik09,padmanabhan10rpp046901,
verlinde11jhep29, 
penrose14fp557, 
giddings15jhep1,
verlinde17spp016,
jacobson16prl201101,
cao-17prd024031,
giddings19fp177, 
cotler-1911.12358}.

%
%

\section{Remarks on thermalization}
\label{thermalization}

Finally, I wish to return to the question of the origin of structures I mentioned in the introduction.  On first sight, the emerging picture reproduces only the physical theories we are familiar with---\emph{quantum mechanics for microscopic degrees of freedom which are disentangled from the \EMO , statistical mechanics for microscopic degrees of freedom which are entangled with the \EMO , and classical mechanics for macroscopic degrees of freedom.}  These are exactly the theories we have been teaching for generations, and which have never provided a promising starting point for investigations of the origin of life.  In systems which thermalize,
entanglement entropy increases, while a local reduction is required for structures to develop.  

There are, however, important exceptions to thermalization, and hence to the applicability of statistical descriptions.  Most prominent among them are integrable systems in general and systems subject to many body localization (MBL)\cite{nandkishoredoi-15arcmp15,abanin-19rmp021001} in particular, where integrability emerges locally\cite{serbyn-13prl127201,huse-14prl174202,imbrie16jsp998}.
More generally, the ETH provides us with a concise framework to investigate when thermalization fails, and hence guides us in our search for promising directions to pursue in our quest to understand the origin of life.

Another observation of potential relevance 
is that thermalization is generally associated with degrees of freedom which are part of the \EMO .  Consider a microscopic degree of freedom, which is not entangled with the \EMO , but entangled with other microscopic degrees of freedom.  If we subject this degree of freedom to a measurement, possibly through a biological process, and assume that a collapse in the sense described above takes place, it will end up in a state which is disentangled from both the other microscopic degrees of freedom and the \EMO . 
(To illustrate this point, note that the collapse will single out one of the terms on the r.h.s.\ of \eqref{eq:martin_epr_fm1}.  So spin 1, which was entangled with spin 2 before the measurement, will be disentangled from all other degrees of freedom afterwards.)
The measurement process hence reduces the entanglement entropy of the degree of freedom we measure.

\section{Summary}
\label{summary}

Let me conclude with a summary of the emergent picture.  We assume that the fundamental theory is a quantum theory, with its evolution given at least approximately by Schrödinger's equation and its relativistic generalizations.  The classical reality we perceive and describe by classical theories is given by the ensemble of macroscopic objects, here abbreviated as the \EMO . Due to interlinking, the wave function of all these objects cannot be factorized.  Therefore we can observe quantum phenomena, and in particular interference phenomena, only for degrees of freedom which are disentangled from the \EMO .  A measurement occurs whenever a microscopic degree of freedom becomes entangled with its environment and hence interlinked with the \EMO .  Then it can no longer be described by an individual wave function, but only through the wave function of the \EMO , which includes the visible universe.  Even though we lack a microscopic theory how a collapse onto one particular branch of the naively emerging multitude of universes occurs, it is reasonable to assume that it does. 

In its fundamental assumptions, the theory is very similar to MWIs, as it only assumes a quantum theory (\AiP ).  The classical reality and the statistical description of our experience are emerging. The difference to previous theories is that we 
introduce interlinking, abandon the von Neumann chain, interprete all entropy as entanglement entropy, and take into account that the length and energy scales relevant for the non-linearities we conjecture to describe the collapse of wave functions are presently inaccessible to us.  

While the fundamental assumptions resemble MWIs more than anything else, the phenomenology we obtain is a refined version of the Copenhagen interpretation.  The key difference is that we do not embed the quantum theory into a classical domain, but find the classical domain within the quantum theory.  

\section*{Acknowledgements}
I wish to thank Dima Abanin, John Cardy, Jens Eisert, Tobias Helbig, Tobias Hofmann, Christoph Holzhey, Frank Wilczek and especially Mark Srednicki for inspiring discussions of parts of this work.  This work was supported by the Deutsche Forschungsgemeinschaft (DFG, German Research Foundation)---Project-ID 258499086---SFB 1170, through the Würzburg-Dresden Cluster of Excellence on Complexity and Topology in Quantum Matter---\textit{ct.qmat} Project-ID 390858490---EXC 2147, and in part through US National Science Foundation Grant No.\ NSF PHY-1748958 while at KITP. 

%

\bibliographystyle{../../../bib/SciPost_bibstyle}
\bibliography{../../../bib/q,../../../bib/stat,../../../bib/gravity,../../../bib/book}

\end{document}